\documentclass[12pt,preprint]{aastex}

\slugcomment{To be published in ApJ}

\shorttitle{Savin et al.} \shortauthors{Rate Coefficients for 
${\rm D}(1s) + {\rm H}^+ \leftrightarrows {\rm D}^+ + {\rm H}(1s)$}

\newcommand{\ionmatha}[2]{{\rm #1 \: {\scriptstyle #2}}}
\newcommand{\ionmathb}[2]{{\rm #1 \: {\scriptscriptstyle #2}}}

\begin{document}

\title{Rate Coefficients for ${\rm D}(1s) + {\rm H}^+
\leftrightarrows {\rm D}^+ + {\rm H}(1s)$ Charge Transfer
and Some Astrophysical Implications}

\author{Daniel Wolf Savin}
\affil{Columbia Astrophysics Laboratory, Columbia University, New
York, NY 10027, USA}
\email{savin@astro.columbia.edu}

\begin{abstract} 

We have calculated the rate coefficients for ${\rm D}(1s) + {\rm H}^+
\leftrightarrows {\rm D}^+ + {\rm H}(1s)$ using recently published
theoretical cross sections.  We present results for temperatures $T$
from 1~K to $2 \times 10^5$~K and provide fits to our data for use in
plasma modeling.  Our calculations are in good agreement with
previously published rate coefficients for $25 \le T \le 300$~K, which
covers most of the limited range for which those results were given.
Our new rate coefficients for $T \gtrsim 100$~K are significantly
larger than the values most commonly used for modeling the chemistry of
the early universe and of molecular clouds.  This may have important
implications for the predicted HD abundance in these environments.
Using our results, we have modeled the ionization balance in high
redshift QSO absorbers.  We find that the new rate coefficients
decrease the inferred D/H ratio by $\lesssim 0.4\%$.  This is a factor
of $\gtrsim 25$ smaller than the current $\gtrsim 10$\% uncertainties
in QSO absorber D/H measurements.

\end{abstract}


\section{Introduction}
\label{sec:Intro}

Deuterium plays an important role in addressing several fundamental
questions in astrophysics.  The deuterium abundance is a key constraint
for models of big bang nucleosynthesis.  Primordial D/H measurements
provide the most sensitive probe of the baryon-to-photon density ratio
$\eta$.  This, in combination with the cosmic microwave background
measurement of the photon density, can be used to determine the
cosmological baryon density \citep{Burl98a,Lemo99a,Tytl00a}.  Deuterium
may also be important in the formation of structure in the 
post-recombination era of the early
universe.  HD is the second most abundant primordial molecule, after
H$_2$, and cooling radiation from it may play a role in the formation
of the first collapsing objects (Puy et al.\ 1993; Stancil, Lepp, \&
Dalgarno 1998).  Lastly, as the universe evolves, deuterium is slowly
destroyed in stars where it is burned into $^3$He.  Mapping the
temporal and spatial variations in the D/H abundance ratio can shed
light on the time history of star formation in different regions of the
cosmos \citep{Tosi98,Tosi98a,Lemo99a}.

Investigations into these issues are carried out through studies of gas
phase D and deuterated molecules in, for example, the early universe
\citep{Gall98a,Stan98a}, QSO absorption systems \citep{OMea01a},
molecular clouds \citep{Tiel92a,Bert99a,Wrig99a} and the interstellar
medium \citep{Lins95a}.  \citet{Stan98a} have given a recent listing of
deuterium models for the early universe, molecular clouds, and the
interstellar medium.  Recent reviews of deuterium observations have
been given by \citet{Lemo99a} and \citet{Tytl00a}.

Interpreting these studies requires an accurate knowledge of all
collision processes involving D.  Particularly significant is the
near-resonant charge transfer (CT) process
\begin{equation}
{\rm D}(1s) + {\rm H}^+ \rightarrow {\rm D}^+ +{\rm H}(1s)
\label{eq:H+D}
\end{equation}
and the reverse process
\begin{equation}
{\rm D}^+ + {\rm H}(1s) \rightarrow {\rm D}(1s) + {\rm H}^+
\label{eq:HD+}
\end{equation}
In the early universe these are two of the most important processes
involving deuterium \citep{Gall98a}.  In molecular clouds,
Process~\ref{eq:H+D} followed by the exothermic reaction
\begin{equation}
{\rm H}_2 + {\rm D}^+ \rightarrow {\rm HD} + {\rm H}^+
\label{eq:H2D+}
\end{equation}
is a major source of HD (Black \& Dalgarno 1973; Dalgarno, Black, \&
Weisheit 1973; Watson 1973).  This is also an important source of HD
in the early universe \citep{Stan98a}.

Recently \citet{Igar99a} and Zhao, Igarashi, \& Lin (2000) have carried
out cross section calculations for Reaction~\ref{eq:H+D} ($\sigma_1$)
and Reaction~\ref{eq:HD+} ($\sigma_2$) using a hyperspherical
close-coupling method.  This technique is free from the ambiguities
associated with the conventional Born-Oppenheimer approach.  Here we
use their results to produce CT rate coefficients for
Reactions~\ref{eq:H+D} ($\alpha_1$) and \ref{eq:HD+}
($\alpha_2$).  In \S~\ref{sec:Rates} we describe how we evaluated
$\alpha_1$ and $\alpha_2$.  Our results are presented in
\S~\ref{sec:Results} and compared with previously published
calculations.  Some astrophysical implications are discussed in
\S~\ref{sec:Implications}.

\section{Calculation of the Rate Coefficients}
\label{sec:Rates}

We consider only capture from and into the $1s$ level of H and D.
The reason for this is twofold.  First, in the sources discussed in
\S~\ref{sec:Intro}, neutral H and D are expected to be found
essentially only in their ground state.  Second, at the
low temperatures (i.e., low collision energies) relevant for these sources
($T \lesssim 3 \times 10^4$~K [i.e., $k_BT \lesssim 3$~eV]), CT into
the $1s$ level is predicted to be over 4 orders of magnitude greater
than capture into other levels.  CT into higher levels does not become
important until collision energies of $\gtrsim 10^3$~eV \citep{Alt94a}.

We use the results of \citet{Igar99a} and \citet{Zhao00a} for
$\sigma_1(E)$ and $\sigma_2(E)$ at center-of-mass energies $E$ from
$2.721 \times 10^{-8}$ to 2.721~eV \citep{Igar01}.  Due to the
binding energy of D$(1s)$ being slighlty larger than that of H$(1s)$,
Reaction~\ref{eq:H+D} is endothermic with a threshold of 3.7~meV
(43~K).  Hence, $\sigma_1$ is predicted to be smaller than $\sigma_2$
at all energies, but particularly at low energies.  As $E$ increases,
$\sigma_1$ and $\sigma_2$ converge, and for $E \gtrsim 2.72$~eV, the
two are predicted to lie within $\lesssim 0.1\%$ of one another
\citep{Zhao00a,Igar01}.  The energy dependences for $\sigma_1$ and
$\sigma_2$ allow us to extend the results of Igarashi and collaborators
to higher energies.  We do this using the calculated cross sections of
\citet{Dalg53a} for the related reaction
\begin{equation}
{\rm H}^+ + {\rm H}(1s) \rightarrow {\rm H}(1s) + {\rm H}^+.
\label{eq:H+H}
\end{equation}
First, to extend the data for $\sigma_2$, we multiply the energy scale
of Dalgarno \& Yadav by $\mu_{HD}/\mu_{HH}$ where $\mu$ is the reduced
mass for the HD and HH systems.  This effectively matches the velocity
scale for each data set.  Then, we multiply the results of Dalgarno \&
Yadav by a factor of 0.959 to set it equal to the results of Igarashi
and collaborators at 1.333~eV.  Next, for energies between 1 and
$\approx 2.72$~eV, we fit the ratio of $\sigma_1/\sigma_2$ from
Igarashi to the formula
\begin{equation}
{\sigma_1 \over \sigma_2} = 1 - {A \over E} - {B \over E^2}
\label{eq:s1vs2}
\end{equation}
which yields $A=2.692\times10^{-3}$ and $B=7.936\times10^{-4}$.
We calculate $\sigma_1$ for energies above
$\approx 2.72$~eV using the scaled cross sections of Dalgarno \& Yadav
multiplied by Equation~\ref{eq:s1vs2}.  

We use the resulting data for $\sigma_1$ and $\sigma_2$ from energies
of $2.721 \times 10^{-8}$ to $10^3$~eV to evaluate the rate
coefficients $\alpha_1(T)$ and $\alpha_2(T)$ as a function of the gas
temperature $T$.  Rate coefficients are calculated numerically using
the desired cross section times the relative velocity and convolving
these results with the appropriate Maxwellian distribution (taking the
reduced mass into account).  Cross sections for energies not calculated
by Igarashi and collaborators or by Dalgarno \& Yadav are evaluated
using a spline interpolation method \citep{Pres92a} for $\sigma(E)$
versus $\log(E)$.

\section{Results and Comparisons}
\label{sec:Results}

\begin{figure}
\plotone{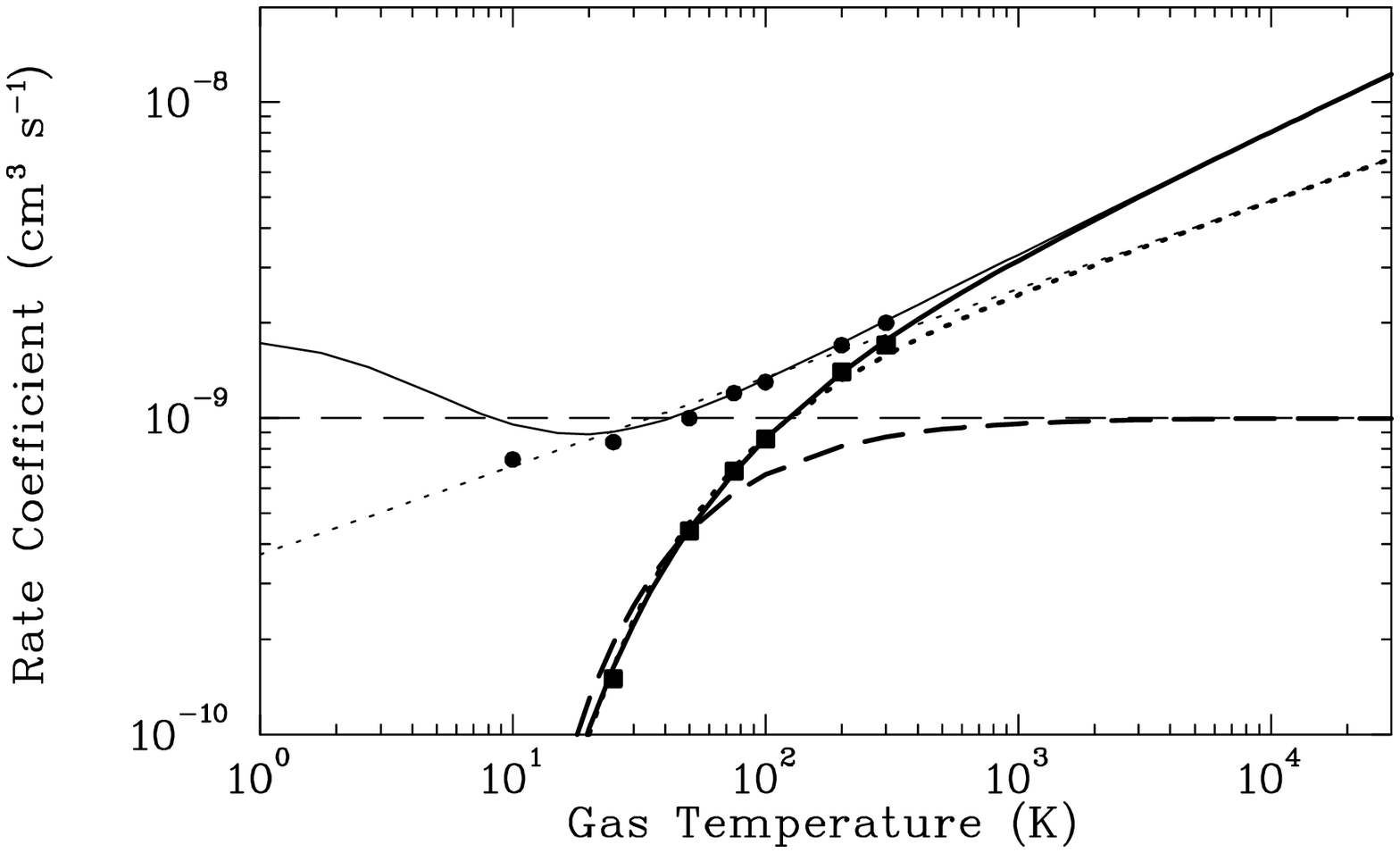}
\caption{Calculated rate coefficients for $\alpha_1(T)$ 
[${\rm D}(1s) + {\rm H}^+ \rightarrow {\rm D}^+ + {\rm H}(1s)$]
and the reverse process $\alpha_2(T)$
[${\rm D}^+ + {\rm H}(1s) \rightarrow {\rm D}(1s) + {\rm H}^+$]
versus
gas temperature $T$.  Results for $\alpha_1$ are from this work ({\it
thick solid curve}), Watson (1976; {\it thick dashed curve}), Watson et
al.\ (1978; {\it squares}), and Galli \& Palla (1998; {\it thick dotted
curve}).  Results for $\alpha_2$ are from this work ({\it thin solid
curve}), Watson (1976; {\it thin dashed curve}), Watson et al.\ (1978;
{\it circles}), and Galli \& Palla (1998; {\it thin dotted curve}).}
\label{fig:rates}
\end{figure}

Our calculated results for $\alpha_1(T)$ and $\alpha_2(T)$ are given in
Tab.~\ref{tab:rates} from 1~K up to $2 \times 10^5$~K.  These results
are also plotted in Fig.~\ref{fig:rates} from 1 to 30,000~K.  As
expected the rate coefficient for the endothermic Reaction~\ref{eq:H+D}
decreases dramatically for $T \lesssim 10$~K.  The rate coefficient for
Reaction~\ref{eq:HD+} decreases slowly with decreasing temperature down
to $T \approx 15$~K.  Below this $\alpha_2$ begins to increase with
decreasing temperature.  We attribute this to the rapid increase in
$\sigma_2$ with decreasing collision energy 
\citep [see Fig.~2 of][]{Igar99a}.

We have fitted our calculated CT rate coefficients using the formula
\begin{equation}
\alpha(T) =  a T^b \exp(-c/T)+d T^e.
\end{equation}
The best fit values are listed in Tab.~\ref{tab:fitparameters}.  The
fits for $\alpha_1$ and $\alpha_2$ are accurate to better than 6\% and
4\%, respectively, for $2.7~{\rm K} \le T \le 2 \times 10^5~{\rm K}$.

Several other groups have carried out detailed calculations for
$\sigma_1$ and $\sigma_2$.  \citet{Davi78a} published results for
$\sigma_1$ from 3.7 to $\approx 100$~meV.  Results for $\sigma_1$ and
$\sigma_2$ were reported by \citet{Hunt77b} for energies from $10^{-3}$
to 7.5~eV, by \citet{Hodg93a} from $10^{-3}$ to 10~eV, and by
\citet{Esry00a} from $\approx 3.7$~meV to 8~eV.  In general the
calculations of Igarashi and collaborators are in good to excellent
agreement with these published results.  The most significant
difference is for energies above $10^{-3}$~eV where the results of
Hunter \& Kuriyan can fall as much as 15\% below those of Igarashi and
collaborators \citep{Igar99a,Zhao00a,Igar01}.  This is partially due to
the accidental overlap of minima in the oscillating cross sections with
the energy points published by \citet{Hunt77b}.

There have been a couple of experimental measurements of $\sigma_1$.
We are unaware of any experimental results for $\sigma_2$.  Absolute
measurements of $\sigma_1$ have been carried out by \citet{Newm82a} for
energies between $\approx 0.1$ and 10~eV.  The theoretical results of
\cite{Zhao00a} and \citet{Esry00a} are in good agreement with these
measurements.  Relative measurements for $\sigma_1$ have carried out by
\citet{Well01a} for energies between threshold and 1~eV.  Good
agreement was found with the calculations of \citet{Esry00a} between
$\approx 0.02$ and 1~eV.  Uncertainties in background subtraction limit
the reliability of the experimental results below 0.02~eV.

Using the results of \citet{Hunt77b}, \citet{Wats78a} calculated
$\alpha_1$ and $\alpha_2$ for a number of temperatures between 10 and
300~K.  These results are listed in Tab.~\ref{tab:rates} and also
plotted in Fig.~\ref{fig:rates}.  For $T \ge 50$~K, the results of
Watson et al.\ agree with ours to better than 5\%.  At 25~K their
results differ from ours by $\approx 8\%$ and at 10~K by $\approx
30\%$.  The differences for $T \le 25$~K are most likely due to the 
uncertainty associated with extrapolating the results of Hunter \& Kuriyan 
to energies below those published \citep{Wats78a}.

\citet{Gall98a} fit the results of \citet{Wats78a} for $\alpha_1$ and
$\alpha_2$.  The resulting fitted rate coefficients are plotted in
Fig.~\ref{fig:rates}.  Between 10 and 300~K, these
fitted rate coefficients agree with our results not quite as well as the
results of Watson et al.  Agreement with our results becomes
progessively worse the further one extrapolates these fitted rates
outside this temperature range.

\citet{Wats76} presented an estimate for $\alpha_1$ and $\alpha_2$
which we plot in Fig.~\ref{fig:rates}.  These estimated rate
coefficients are in poor agreement with our results here, differing
significantly in both the values and temperature dependences of
$\alpha_1$ and $\alpha_2$.

\section{Some Astrophysical Implications}
\label{sec:Implications}

\subsection{The Early Universe}

Recently, results from a number of different chemical models of the
early universe have been published.  For these models, \citet{Puy93a}
and \citet{Stan98a} used the estimated rate coefficients of
\citet{Wats76}.  \citet{Gall98a} used their fits to the results of
\citet{Wats78a}.  For redshifts $z \gtrsim 50$, where the gas
temperature is predicted to be $\gtrsim 50$~K \citep{Puy93a}, the rate
coefficients used by Puy et al.\ and Stancil et al.\ begin to differ
significantly from our newly calculated results.  At $z\approx 400$
\citep[$T\approx 1000$~K;][]{Puy93a}, the rate coefficients used by
them are a factor of $\approx 3$ smaller than our results.  In
contrast, the extrapolated rate coefficients used by Galli \& Palla are
only $\approx 22\%$ smaller.  Determining the full implications of our
new rate coefficients will require re-running updated versions of these
various chemical models of the early universe.

\subsection{Molecular Clouds}

Modeling studies of molecular clouds have been carried out recently by
Millar, Bennett, \& Herbst (1989), Pineau des For\^ets, Roueff, \&
Flower (1989), Heiles, McCullough, \& Glassgold (1993),
\citet{Rodg96a}, and \citet{Timm96}.  These
studies have all used the results of \citet{Wats76} for $\alpha_1$ and
$\alpha_2$ and hence significantly underestimate these two rate
coefficients for $T\gtrsim 100$~K.  Because Reaction~\ref{eq:H+D}
followed by Reaction~\ref{eq:H2D+} is predicted to be a major source of
HD in molecular clouds (Black \& Dalgarno 1973; Dalgarno, Black, \&
Weisheit 1973; Watson 1973), underestimating $\alpha_1$ could in turn
lead to an underestimate in the amount of HD produced in these clouds.

\subsection{High Redshift QSO Absorption Systems}

Observations of high redshift QSO absorption systems are used to infer
the primordial D/H ratio.  These studies are carried out assuming that
the \ion{D}{1}/\ion{H}{1} ratio is identical to that of D/H
\citep{Burl98a}.  Here we investigate the validity of this assumption
in light of the different values for $\alpha_1$ and $\alpha_2$.

The D/H ratio inferred from these observations is given by
\begin{equation}
{n_{\rm D} \over n_{\rm H}} =  
{f_{\ionmathb{H}{I}} \over f_{\ionmathb{D}{I}}}
{N(\ionmatha{D}{I}) \over N(\ionmatha{H}{I})}
\label{eq:nDvnH}
\end{equation}
where $n_{\rm D}$ is number density of D, $N(\ionmatha{D}{I})$ is the
column density of \ion{D}{1}, and $f_{\ionmathb{D}{I}}$ is the
abundance of \ion{D}{1} relative to the total abundance of D.  Similar
definitions exist for H and \ion{H}{1}.  We can write
$f_{\ionmathb{D}{I}}$ as
\begin{equation}
{1 \over f_{\ionmathb{D}{I}}} 
= 
1 + {n_{\ionmathb{D}{II}} \over n_{\ionmathb{D}{I}}}.
\end{equation}
A similar expression can be written for $f_{\ionmathb{H}{I}}$. 

Currently there are believed to be six reliable measurements of D/H in
high redshift QSO absorbers \citep{Pett01a}. 
These measurements all assume 
$f_{\ionmathb{D}{I}}=f_{\ionmathb{H}{I}}$.  To determine
the validity of this assumption, we evaluate
\begin{equation}
{f_{\ionmathb{H}{I}} \over f_{\ionmathb{D}{I}}}
=
{1 + {n_{\ionmathb{D}{II}} / n_{\ionmathb{D}{I}}} \over
 1 + {n_{\ionmathb{H}{II}} / n_{\ionmathb{H}{I}}}}
\label{eq:fH1vfD1}
\end{equation}
using our new results for $\alpha_1$ and $\alpha_2$.

At the inferred temperatures in these six absorbers ($T\approx 1.1
\times 10^4$~K), the gas is predicted to have an insignificant
abundance of molecules (Petitjean, Srianand, \& Ledoux 2001).  The
ionization balance of D in these QSO absorbers can therefore be written
\begin{equation}
{n_{\ionmathb{D}{II}} \over n_{\ionmathb{D}{I}}}
=
{
\beta_{\ionmathb{D}{I}} + 
n_{\rm e} C_{\ionmathb{D}{I}} +
n_{\ionmathb{H}{II}} \alpha_{{\rm D} + {\rm H}^+} +
\sum_{{\rm X}^{q+}} n_{{\rm X}^{q+}} \alpha_{{\rm D} + {\rm X}^{q+}}
\over
n_{\rm e} R_{\ionmathb{D}{II}} +
n_{\ionmathb{H}{I}}  \alpha_{{\rm H} + {\rm D}^+} +
\sum_{\rm X} n_{\rm X} \alpha_{{\rm X} + {\rm D}^+}
}.
\label{eq:Dbalance}
\end{equation}
In the numerator on the right-hand-side (RHS) of this equation,
$\beta_{\ionmathb{D}{I}}$ is the photoionization (PI) rate of
\ion{D}{1} due to the radiation field and accounts for further
ionization due to the resulting non-thermal photoelectrons, $n_{\rm e}$
is the electron density, $C_{\ionmathb{D}{I}}$ is the electron impact
ionization (EII) rate coefficient due to thermal electrons,
$\alpha_{{\rm D} + {\rm H}^+}$ is the rate coefficient for ${\rm D} +
{\rm H}^+$ collisions producing D$^+$, and similarly for $\alpha_{{\rm
D} + {\rm X}^{q+}}$ where X$^{q+}$ represents a $q$-times charged ion
of element X and the sum over X$^{q+}$ includes the ions of all
elements except for those of H and D.  Here, $\alpha_{{\rm D} + {\rm
H}^+}$ and $\alpha_{{\rm D} + {\rm X}^{q+}}$ are purely CT rate
coefficients.  This is because collisions which leave both colliding
particles in an ionized state are predicted to be insignificant at the
temperatures of interest (Janev, Presnyakov, \& Shevelko 1985).
In the denominator on the RHS, $R_{\ionmathb{D}{II}}$ is the radiative
recombination (RR) rate coefficient for \ion{D}{2}, $\alpha_{{\rm H} +
{\rm D}^+}$ is the CT rate coefficient for ${\rm H} + {\rm D}^+$
collisions producing D, and similarly for $\alpha_{{\rm X} + {\rm
D}^+}$.

The PI rates and EII and RR rate coefficients for D and H are expected
to be essentially identical \citep{Gall98a,Stan98a}.  The differences
in the energy level structure of D and H have an insignificant effect
on these processes.  Thus in Eq.~\ref{eq:Dbalance}, we can substitute
\begin{eqnarray}
\beta_{\ionmathb{H}{I}} = \beta_{\ionmathb{D}{I}}, \\
C_{\ionmathb{H}{I}} = C_{\ionmathb{D}{I}},
\end{eqnarray}
and
\begin{eqnarray}
R_{\ionmathb{H}{II}} = R_{\ionmathb{D}{II}}.
\end{eqnarray}
Next we add and subtract
\begin{equation}
n_{\ionmathb{D}{II}} \alpha_{{\rm H} + {\rm D}^+} +
\sum_{{\rm X}^{q+}} n_{{\rm X}^{q+}} \alpha_{{\rm H} + {\rm X}^{q+}}
\end{equation}
to the numerator on the RHS of Eq.~\ref{eq:Dbalance} and
\begin{equation}
n_{\ionmathb{D}{I}}  \alpha_{{\rm D} + {\rm H}^+} +
\sum_{\rm X} n_{\rm X} \alpha_{{\rm X} + {\rm H}^+}
\end{equation} 
to the denominator.  We note that using Eq.~\ref{eq:Dbalance}
we get $n_{\ionmathb{H}{II}} /
n_{\ionmathb{H}{I}}$ by interchanging all charge states of D with the
corresponding charge states of H (and vice versa).  Hence, we can
rewrite Eq.~\ref{eq:Dbalance} as
\begin{equation}
{n_{\ionmathb{D}{II}} \over n_{\ionmathb{D}{I}}}
=
{n_{\ionmathb{H}{II}} \over n_{\ionmathb{H}{I}}}
\Biggl(
{1 + \gamma_1/\delta \over 1 + \gamma_2/\delta}
\Biggr)
\label{eq:Dbalance2}
\end{equation}
where
\begin{equation}
\gamma_1 =
\alpha_{{\rm D} + {\rm H}^+} 
- 
{n_{\ionmathb{D}{II}} \over n_{\ionmathb{H}{II}}} 
\alpha_{{\rm H} + {\rm D}^+} 
+
\sum_{{\rm X}^{q+}}
{n_{{\rm X}^{q+}} \over n_{\ionmathb{H}{II}}}
\Biggl( 
\alpha_{{\rm D} + {\rm X}^{q+}}
-
\alpha_{{\rm H} + {\rm X}^{q+}}
\Biggr),
\label{eq:gamma1}
\end{equation}
\begin{equation}
\gamma_2 =
\alpha_{{\rm H} + {\rm D}^+} 
- 
{n_{\ionmathb{D}{I}} \over n_{\ionmathb{H}{I}}} 
\alpha_{{\rm D} + {\rm H}^+} 
+
\sum_{\rm X}
{n_{\rm X} \over n_{\ionmathb{H}{I}}}
\Biggl( 
\alpha_{{\rm X} + {\rm D}^+}
-
\alpha_{{\rm X} + {\rm H}^+}
\Biggr),
\label{eq:gamma2}
\end{equation}
and 
\begin{equation}
\delta =
{n_{\rm e} \over n_{\ionmathb{H}{I}}}
R_{\ionmathb{H}{II}}
+
{n_{\ionmathb{D}{II}} \over n_{\ionmathb{H}{I}}}
\alpha_{{\rm D} + {\rm H}^+}
+
\sum_{\rm X} 
{n_{\rm X} \over n_{\ionmathb{H}{I}}}
\alpha_{{\rm X} + {\rm H}^+}.
\label{eq:delta}
\end{equation}

We can simplify $\gamma_1$ and $\gamma_2$. First we note that
$\alpha_{{\rm D} + {\rm H}^+}=\alpha_1$ and $\alpha_{{\rm H} + {\rm
D}^+}=\alpha_2$; and at the temperatures of interest $\alpha_1 \approx
\alpha_2 \approx 8.3 \times 10^{-9}$~cm$^3$~s$^{-1}$.  Now, to a first
approximation $n_{\ionmathb{D}{II}}/n_{\ionmathb{H}{II}}$ and
$n_{\ionmathb{D}{I}}/n_{\ionmathb{H}{I}}$ will be equal to the
primordial D/H value which we take to be $\approx 2 \times 10^{-5}$
\citep[from][]{Pett01a}.  Hence the second term in Eqs.~\ref{eq:gamma1}
and \ref{eq:gamma2} is roughly 5 orders of magnitude smaller than the
first term and can be dropped.

At energies important for $T \approx 1.1 \times 10^4$~K, we note that
$\sigma_1 \approx \sigma_2$.  Similarly, we expect at these
temperatures $\sigma_{{\rm D} + {\rm X}^{q+}}(v) \approx \sigma_{{\rm
H} + {\rm X}^{q+}}(v)$ and $\sigma_{{\rm X} + {\rm D}^+}(v) \approx
\sigma_{{\rm X} + {\rm H}^+}(v)$, where $v$ is the relative velocity.
As a result, we have $\alpha_{{\rm D} + {\rm X}^{q+}}\sqrt{\mu_{DX}}
\approx \alpha_{{\rm H} + {\rm X}^{q+}}\sqrt{\mu_{HX}}$ and
$\alpha_{{\rm X} + {\rm D}^+}\sqrt{\mu_{DX}} \approx \alpha_{{\rm X} +
{\rm H}^+}\sqrt{\mu_{HX}}$.  Here, $\mu_{HX} \approx 1$ and $\mu_{DX}
\approx 2$ are the reduced masses.  For those ions where CT is
important in photoionized plasmas (e.g., QSO absorbers) we estimate
that $\alpha_{{\rm D} + {\rm X}^{q+}}$, $\alpha_{{\rm H} + {\rm
X}^{q+}}$, $\alpha_{{\rm X} + {\rm D}^+}$, and $\alpha_{{\rm X} + {\rm
H}^+}$ will all be $\lesssim 10^{-9}$~cm$^3$~s$^{-1}$ \citep{King96a}.
The expressions in the parenthesis in the third term in
Eqs.~\ref{eq:gamma1} and \ref{eq:gamma2} are thus $\lesssim 4 \times
10^{-10}$~cm$^3$~s$^{-1}$.  Furthermore, we note that the metallicity
in these absorbing systems are $\approx 10^{-2}$ solar \citep{Pett01a}
and that we expect $n_{{\rm X}^{q+}}/n_{\ionmathb{H}{II}}$ and $n_{\rm
X}/n_{\ionmathb{H}{I}}$ will be within a couple of orders of magnitude
of these reduced abundances.  As a result, we can also drop the third
term in Eqs.~\ref{eq:gamma1} and \ref{eq:gamma2}.  With these
approximations we can rewrite Eq.~\ref{eq:Dbalance2} as
\begin{equation}
{n_{\ionmathb{D}{II}} \over n_{\ionmathb{D}{I}}}
\approx
{n_{\ionmathb{H}{II}} \over n_{\ionmathb{H}{I}}}
\Biggl(
{\alpha_1 \over \alpha_2}
\Biggr)
\Biggl(
{1 + \delta/\alpha_1 \over 1 + \delta/\alpha_2}
\Biggr).
\label{eq:Dbalance3}
\end{equation}
As $\alpha_1$ and $\alpha_2$ differ by $\lesssim 0.4\%$ at 
$T \approx 1.1 \times 10^4$~K, Eq.~\ref{eq:Dbalance3} reduces to
\begin{equation}
{n_{\ionmathb{D}{II}} \over n_{\ionmathb{D}{I}}}
\approx
{n_{\ionmathb{H}{II}} \over n_{\ionmathb{H}{I}}}
\Biggl(
{\alpha_1\over \alpha_2}
\Biggr).
\label{eq:Dbalance4}
\end{equation}
Because $\sigma_1$ has a threshold, in contrast with $\sigma_2$, the
ratio $\alpha_1/\alpha_2$ will always be less than 1.  For the
temperature of interest here, we have
\begin{equation}
{n_{\ionmathb{D}{II}} \over n_{\ionmathb{D}{I}}}
\approx
0.996{n_{\ionmathb{H}{II}} \over n_{\ionmathb{H}{I}}}.
\end{equation}
Substituting this results into Eqs.~\ref{eq:nDvnH} and \ref{eq:fH1vfD1},
we find
\begin{eqnarray}
{n_{\rm D} \over n_{\rm H} }
= \left\{ 
\begin{array}{ll}
\approx 0.996N(\ionmatha{D}{I})/N(\ionmatha{H}{I})
& \ \ \ \ \ \ \ (n_{\ionmathb{H}{II}} \gg n_{\ionmathb{H}{I}}) \\
\ \ \ 1.000 N(\ionmatha{D}{I})/N(\ionmatha{H}{I})
& \ \ \ \ \ \ \ (n_{\ionmathb{H}{II}} \ll n_{\ionmathb{H}{I}}).
\end{array}
\right.
\end{eqnarray}
The uncertainty in the above factor of 0.996 depends on the accuracy of
the theoretical cross sections we have used here.  Verifying the
accuracy of these cross sections will require further theoretical and
experimental studies.  However, we note that the variation of $\approx
0.4\%$ in $n_{\rm D}/n_{\rm H}$ is a factor of $\gtrsim 25$ smaller
than the current $\gtrsim 10\%$ uncertainties in QSO absorber D/H
measurements.  Thus, it is likely to be some time before the subtle
differences in Reactions~\ref{eq:H+D} and \ref{eq:HD+} become important
for primordial D/H measurements.

\acknowledgements

We would like to thank E. Behar, A. Igarashi, K. Korista, A. Lidz, and
P. A. Stancil for stimulating discussions.  We also wish to thank A.
Igarashi and H. Sadeghpour for kindly providing their results in
electronic format.  This work was supported in part by NASA Space
Astrophysics Research and Analysis Program grant NAG5-5261.

\vfill
\eject

\vfill
\eject

\begin{deluxetable}{cccccc}
\tablecaption{Calculated rate coefficients for $\alpha_1(T)$ 
[${\rm D}(1s) + {\rm H}^+ \rightarrow {\rm D}^+ + {\rm H}(1s)$]
and the reverse process $\alpha_2(T)$
[${\rm D}^+ + {\rm H}(1s) \rightarrow {\rm D}(1s) + {\rm H}^+$]
versus gas temperature $T$.
\label{tab:rates}}
\tablewidth{0pt}
\tablehead{
\colhead{T(K)} 
& \multicolumn{2}{c}{$\alpha_1$ (cm$^3$ s$^{-1}$)}
& \colhead{}
& \multicolumn{2}{c}{$\alpha_2$ (cm$^3$ s$^{-1}$)} \\
\cline{2-3}
\cline{5-6}
\colhead{}
& \colhead{Present}
& \colhead{Watson et al.\tablenotemark{a}}
& \colhead{}
& \colhead{Present}
& \colhead{Watson et al.\tablenotemark{a}}
}
\startdata
1      & 3.88E-28 & & & 1.73E-09 & \\
2.7    & 1.79E-16 & & & 1.45E-09 & \\
5      & 2.21E-13 & & & 1.18E-09 & \\
10     & 1.30E-11 & 1.0E-11 & & 9.55E-10 & 7.4E-10 \\
15     & 5.12E-11 & & & 8.95E-10 & \\
20     & 1.04E-10 & & & 8.90E-10 & \\
25     & 1.63E-10 & 1.5E-10 & & 9.06E-10 & 8.4E-10 \\
30     & 2.22E-10 & & & 9.31E-10 & \\
35     & 2.81E-10 & & & 9.59E-10 & \\
40     & 3.38E-10 & & & 9.89E-10 & \\
45     & 3.93E-10 & & & 1.02E-09 & \\
50     & 4.45E-10 & 4.4E-10 & & 1.05E-09 & 1.0E-09 \\
75     & 6.75E-10 & 6.8E-10 & & 1.20E-09 & 1.2E-09 \\
100    & 8.63E-10 & 8.6E-10 & & 1.325E-09 & 1.3E-09 \\
200    & 1.40E-09 & 1.4E-09 & & 1.73E-09 & 1.7E-09 \\
300    & 1.76E-09 & 1.7E-09 & & 2.03E-09 & 2.0E-09 \\
400    & 2.05E-09 & & & 2.28E-09 & \\
500    & 2.29E-09 & & & 2.49E-09 & \\
600    & 2.50E-09 & & & 2.68E-09 & \\
700    & 2.68E-09 & & & 2.85E-09 & \\
800    & 2.85E-09 & & & 3.01E-09 & \\
900    & 3.01E-09 & & & 3.15E-09 & \\
1000   & 3.15E-09 & & & 3.29E-09 & \\
1500   & 3.75E-09 & & & 3.86E-09 & \\
2000   & 4.23E-09 & & & 4.32E-09 & \\
3000   & 5.00E-09 & & & 5.07E-09 & \\
4000   & 5.61E-09 & & & 5.67E-09 & \\
5000   & 6.13E-09 & & & 6.18E-09 & \\
6000   & 6.59E-09 & & & 6.63E-09 & \\
7000   & 7.00E-09 & & & 7.04E-09 & \\
8000   & 7.37E-09 & & & 7.41E-09 & \\
9000   & 7.72E-09 & & & 7.75E-09 & \\
10000  & 8.04E-09 & & & 8.07E-09 & \\
11000  & 8.34E-09 & & & 8.37E-09 & \\
12000  & 8.62E-09 & & & 8.65E-09 & \\
13000  & 8.89E-09 & & & 8.92E-09 & \\
14000  & 9.15E-09 & & & 9.17E-09 & \\
15000  & 9.39E-09 & & & 9.42E-09 & \\
20000  & 1.05E-08 & & & 1.05E-08 & \\
25000  & 1.14E-08 & & & 1.14E-08 & \\
30000  & 1.22E-08 & & & 1.23E-08 & \\
35000  & 1.30E-08 & & & 1.30E-08 & \\
40000  & 1.37E-08 & & & 1.37E-08 & \\
50000  & 1.49E-08 & & & 1.49E-08 & \\
75000  & 1.73E-08 & & & 1.74E-08 & \\
100000 & 1.93E-08 & & & 1.93E-08 & \\
200000 & 2.50E-08 & & & 2.50E-08 & \\
\enddata
\tablenotetext{a}{\citet{Wats78a}}
\end{deluxetable}

\vfill
\eject

\begin{deluxetable}{cccccc}
\tablecaption{Fit parameters for
our calculated rate coefficients $\alpha_1(T)$ 
[${\rm D}(1s) + {\rm H}^+ \rightarrow {\rm D}^+ + {\rm H}(1s)$]
and the reverse process $\alpha_2(T)$
[${\rm D}^+ + {\rm H}(1s) \rightarrow {\rm D}(1s) + {\rm H}^+$].
\label{tab:fitparameters}}
\tablewidth{0pt}
\tablehead{
\colhead{Rate Coefficient} &
\colhead{$a$} &
\colhead{$b$} &
\colhead{$c$} &
\colhead{$d$} &
\colhead{$e$} \\
\colhead{}    &
\colhead{(cm$^3$ s$^{-1}$)} &
\colhead{} &
\colhead{(K)} &
\colhead{(cm$^3$ s$^{-1}$)} &
\colhead{} 
}
\startdata
$\alpha_1$ & 2.00E-10 & 0.402 & 37.1 & -3.31E-17 &  1.48 \\
$\alpha_2$ & 2.06E-10 & 0.396 & 33.0 &\ 2.03E-09 & -0.332 \\
\enddata
\end{deluxetable}

\end{document}